\documentclass[aps,twocolumn,prl]{revtex4}
\usepackage[dvips]{graphicx}
\begin{document}
\draft

\title{ Condensate Heating by Atomic Losses }

\author{ Jacek Dziarmaga and Krzysztof Sacha}

\address{
Intytut Fizyki Uniwersytetu Jagiello\'nskiego, 
ul.~Reymonta 4, 30-059 Krak\'ow, Poland 
}

\date{ February 15, 2003 }

\begin{abstract}
Atomic Bose-Einstein condensate is heated by atomic losses. 
Predicted depletion ranges from $1\%$ for a uniform 3D condensate 
to around $10\%$ for a quasi-1D condensate in a harmonic trap. 
\end{abstract}
\maketitle

PACS 03.75-b, 03.75.Gg, 03.75.Hh  


{\bf Heating by atomic losses.---} An ideal Bose-Einstein condensate (BEC)  
is a state where all bosons occupy the same single particle state
$\phi_0$. So far the atomic BEC is the closest to this ideal \cite{Nobel}.
However, even the atomic BEC is not perfect because atoms are depleted
from $\phi_0$ by atom-atom interactions (quantum depletion) and thermal
fluctuations (thermal depletion). Moreover, atomic condensates last only
for tens of seconds before atomic losses empty the trap. This process is
dominated by 3-body losses, where 3 atoms collide to form a bi-atomic
molecule and an atom with large kinetic energy, and then both of them
escape from the trap. There are also 1-body losses,
where individual atoms are kicked out from the trap by external heating
agents, see e.g. Ref.\cite{losses}. While the 1-body losses can, at
least in principle, be minimized, so far the more intrinsic 3-body losses
have not been eliminated.

Atomic losses from a trap can be modeled by repeated action of an
annihilation operator $\hat\psi(\vec x)$. An ideal condensate
$|N:\phi_0\rangle$ is a ``fixed point'' of the annihilation operator:
action of $\hat\psi(\vec x)$ neither depletes an ideal condensate nor
changes its wave function $\phi_0$,
$\hat\psi(\vec x)|N:\phi_0\rangle=
\sqrt{N}\phi_0(\vec x)|N-1:\phi_0\rangle$. 
A destructive measurement of atomic positions in a trap can also be
described by repeated application of $\hat\psi(\vec x)$. In Refs.\cite{JY}
it was shown on a few examples that such a repeated annihilation can
gradually increase a condensed fraction of atoms remaining in a trap.
Annihilation is driving the remaining atoms toward an ideal condensate.
This phenomenon is a foundation for a quite popular believe that ``atomic
losses improve a condensate''. This is not a sound foundation.

In the destructive measurement all annihilations happen at the same time,
or in a very short measurement time. Hamiltonian of atoms has no time to
do anything before all atoms are annihilated. In contrast, in a typical
experiment it is the Hamiltonian that is much faster than atomic losses.  
An ideal condensate $|N:\phi_0\rangle$ is not an eigenstate of the
Hamiltonian: atom-atom interactions are depleting atoms from the
condensate wave function $\phi_0$. On the other hand, eigenstates of the
Hamiltonian $\hat H$ are not ``fixed points'' of the annihilation
operator:  $\hat\psi(\vec x)$ applied to an eigenstate of $\hat H$ gives a
non-stationary state because $\hat\psi(\vec x)$ does not commute with
$\hat H$, $[\hat\psi(\vec x),\hat H]\neq 0$. A competition between the
Hamiltonian depleting a condensate and the atomic losses increasing the
condensed fraction leads to a state that is neither an ideal condensate
nor the $N$-body ground state. A condensate is heated by atomic losses ---
this phenomenon will be more quantified in the following.

In Ref.\cite{Timmermans} Timmermans demonstrated that heating of {\it
fermionic} atoms by atomic losses is a serious obstacle on the way to
atomic superconductors. In the case of {\it fermions} the mechanism is
simple: atomic losses create holes in the Fermi sea. In the case of {\it
bosons} the heating is a conceptually more subtle effect. One has to
realize the interplay between the losses and the Hamiltonian to see that
losses are in fact heating a condensate. However, as we will see below,
a conceptually subtle effect is not necessarily a quantitatively subtle
effect. It is a serious and rather fundamental limitation on
quantum coherence of atomic BEC.


{\bf Master equation.---} The non-unitary (due to atomic losses)  
evolution of trapped atoms is described by a master equation \cite{master}
\begin{equation}
\frac{d\hat\rho}{dt}=
\frac{1}{i\hbar}[\hat H,\hat\rho]+
\sum_{l}\gamma_l \int d^3x~ 
{\cal D}\left[\hat\psi^l(\vec x)\right]\rho\;.
\label{master}
\end{equation}
Here $\hat H$ is a Hamiltonian of trapped atoms
\begin{equation}
\hat H=\int d^3x~
\left[
\frac{\hbar^2}{2m} \nabla\hat\psi^{\dagger}\nabla\hat\psi+
V(\vec x)\hat\psi^{\dagger}\hat\psi+
\frac{g}{2}\hat\psi^{\dagger}\hat\psi^{\dagger}\hat\psi\hat\psi
\right]~,
\label{H}
\end{equation}
with $g=4\pi\hbar^2a/m$, where $a$ is the $s$-wave scattering length.
${\cal D}\left[\hat\psi^l(\vec x)\right]$ is a Lindblad superoperator 
\cite{Lindblad},
\begin{equation}
{\cal D}[\hat a]\rho\equiv
\hat a \rho \hat a^{\dagger} -
\frac12 \hat a^{\dagger} \hat a \rho -
\frac12 \rho \hat a^{\dagger} \hat a~,
\label{Lindblat}
\end{equation}
describing $l$-body losses. 


{\bf Bogoliubov theory.---} We assume that almost all $N$ atoms occupy 
a common condensate wave function $\phi_0(\vec x)$, which solves a stationary 
Gross-Pitaievskii equation \cite{GPE}
\begin{equation}
\mu\phi_0=
-\frac{\hbar^2}{2m}\nabla^2\phi_0+
V(\vec x)\phi_0+
Ng|\phi_0|^2\phi_0~.
\label{GPE}
\end{equation}
The annihilation operator can be split into a condensed part and 
a non-condensed part which is then approximated by an expansion in 
Bogoliubov modes \cite{BT},
\begin{eqnarray}	
\hat\psi(\vec x) &\approx& 
N^{1/2}\phi_0(\vec x)+\delta\hat\psi(\vec x) \approx
\nonumber\\
&&
N^{1/2}\phi_0(\vec x)+
\sum_{m=1}^{\infty} 
\left[
\hat b_m u_m(\vec x) +
\hat b_m^{\dagger} v_m^*(\vec x)
\right]~.
\label{expansion}
\end{eqnarray}
Here $\hat b$'s are bosonic quasiparticle annihilation operators, and
the wave functions $u_m$ and $v_m$ satisfy Bogoliubov-de Gennes equations
\cite{BT}.
 
The operator of a number of atoms depleted from $\phi_0$ is 
$d\hat N=\int d^3x~\delta\hat\psi^{\dagger}\delta\hat\psi$. In the Bogoliubov 
vacuum state $|0_b\rangle$ without any quasiparticles, 
$\hat b_m|0_b\rangle=0$, the number of depleted atoms is
\begin{equation}
dN^{(0)}=
\sum_m \int d^3x~|v_m|^2 \equiv 
\sum_m dN^{(0)}_m.
\label{dN0}
\end{equation}
More generally, in a state with exactly $n_m$ quasiparticles in a mode $m$ the
number of depleted atoms is
\begin{equation}
dN=\sum_m \left[dN_m^{(0)}+\left(1+2dN_m^{(0)}\right)n_m\right]~.
\label{dN}
\end{equation}  
Here we used $\int d^3x~(|u_m|^2-|v_m|^2)=1$.


{\bf Master equation in the quasiparticle representation.---} Bogoliubov
expansion (\ref{expansion}) can be used to rewrite the master equation
(\ref{master}) in the quasiparticle representation. Expansion to second
order in $b$'s in Eqs.(\ref{H},\ref{master}) results in a Bogoliubov
Hamiltonian \cite{BT}, which is a sum of harmonic oscillators $\hat H
\approx \sum_m \hbar\omega_m \hat b_m^{\dagger} \hat b_m$, and an
approximate master equation
\begin{eqnarray}
\frac{d\hat\rho}{dt}&=&
\sum_m
-i\omega_m[\hat b_m^{\dagger}\hat b_m,\hat\rho]+
\label{masterb}\\
&&
\sum_{ml} l \gamma_l \alpha_{lm} N^{l-1}
\left[ (1+n_{lm}) {\cal D}[\hat b_m] \rho +
         n_{lm}   {\cal D}[\hat b_m^{\dagger}] \rho \right]~.
\nonumber
\end{eqnarray}
The coefficients are defined by integrals
\begin{eqnarray}
\int d^3x~ |\phi_0|^{2l-2} |u_m|^2 &=&
\alpha_{lm}(1+n_{lm}) ~, \label{1+n}\\
\int d^3x~ |\phi_0|^{2l-2} |v_m|^2 &=&
\alpha_{lm} n_{lm}    ~. \label{n}
\end{eqnarray}
In addition to small depletion, derivation of the master equation
(\ref{masterb}) requires the rotating wave approximation (RWA).  In the
RWA we neglect all terms of the form $\hat b\hat b$, 
$\hat b^{\dagger}\hat b^{\dagger}$, or $\hat b_m\hat b_n^{\dagger}$
for $m\neq n$, but keep all terms like $\hat b_m\hat b_m^{\dagger}$
or $\hat b_m^{\dagger}\hat b_m$. The RWA is accurate when the
Hamiltonian evolution is much faster than atomic losses, or more precisely
$\omega_m \gg \sum_l l \gamma_l \alpha_{lm} N^{l-1} n_{lm}$.
This condition is satisfied in all present day experiments. 


{\bf A thermal state.---} Due to atomic losses the coefficients in
(\ref{masterb}) are not constant. However, we fix them (for a while) and
analyze a stationary state of the resulting master equation. Later on we
will see that such an analysis allows us to predict a lower bound for a
stationary depletion of a condensate caused by atomic losses.

A remarkable thing is that the master equation (\ref{masterb}), with the
coefficients fixed, can be recognized to describe a set of harmonic
oscillators (numbered by $m$) coupled to external heat reservoirs
(numbered by $l$). Every oscillator relaxes to a thermal state. When
atomic losses are dominated by only one of the channels $l$, then average
numbers of quasiparticles in the thermal states are
\begin{equation}
n_m=
{\rm Tr}~ \hat b^{\dagger}_m \hat b_m~ \hat \rho(t\to\infty)=
n_{lm}~.
\end{equation} 
When many channels $l$ are involved, then the averages $n_m$ can be 
obtained 
from equations
\begin{equation}
\frac{1+n_m}{n_m}=
\frac{\sum_l l \gamma_l \alpha_{lm} N^{l-1} (1+n_{lm})  }
     {\sum_l l \gamma_l \alpha_{lm} N^{l-1}   n_{lm}    }~.
\label{nm}
\end{equation}
Every oscillator $m$ can be assigned to an inverse temperature $\beta_m$ which 
follows from a textbook formula
$n_m=(e^{\beta_m\hbar\omega_m}-1)^{-1}$.
The thermal state is
\begin{equation}
\rho(t\to\infty)=\otimes_m e^{-\beta_m\hbar\omega_m\hat b^{\dagger}_m\hat b_m}.
\label{thermal}
\end{equation}


{\bf A thermal state of a uniform BEC.---} For a 3D condensate with
$\phi_0(\vec x)=\mbox{const}$, the $n_{lm}=dN^{(0)}_m$ are independent of
$l$, compare Eqs.(\ref{1+n},\ref{n},\ref{dN0}). In a uniform condensate of
density $\rho_c$ a phonon of momentum $\hbar k$ has energy \cite{BT}
\begin{equation}
\hbar\omega_k=
\sqrt{\frac{\hbar^2k^2}{2m}\left(\frac{\hbar^2k^2}{2m}+2g\rho_c\right)}
\stackrel{ k^2 \ll \frac{mg\rho_c}{\hbar^2} }{\approx} 
c \hbar k ~.
\label{c}
\end{equation}
Here $c=\sqrt{g\rho_c/m}$ is a velocity of sound. In the thermal state
(\ref{thermal}) there are on average \cite{BT}
\begin{equation}
n_k=\int d^3x~|v_k|^2=
\frac{\frac{\hbar^2k^2}{2m}+g\rho_c }{2\hbar\omega_p}-\frac12
\stackrel{ k^2 \ll \frac{mg\rho_c}{\hbar^2} }{\approx} 
\frac{c}{2\hbar k}
\label{nk}
\end{equation}
phonons of momentum $\hbar k$. With the general formula (\ref{dN}) we can
calculate a fraction of depleted atoms in the thermal state
$d=dN/N=\sqrt{4\pi\rho_c a^3}$. For a typical condensate density of
$\rho_c=10^{20}{\rm m}^{-3}$ we find depletions $d_{^{23}{\rm Na}}=0.44\%$
and $d_{^{87}{\rm Rb}}= 1.55\%$ at the scattering lengths of $a=2.5\;$nm
and $a=5.8\;$nm respectively.

The depletion is dominated by a contribution from small $k$ where the
number of quasiparticles $n_k$ in Eq.(\ref{nk}) is divergent.  A
remarkable thing is that for small $k$ the equipartition of energy
$n_k\hbar\omega_k=\beta^{-1}$ yields the same temperature $T$ for all
phonons. The temperature is $T_{^{23}{\rm Na}}= 76\;{\rm nK}$ and
$T_{^{87}{\rm Rb}}= 37\;{\rm nK}$.

These estimates are valid for a 3D uniform condensate.  In less than 3D
the infrared divergence $n_k\sim k^{-1}$ results in a divergent depleted
fraction $d$.  Anticipating a much larger but finite $d$ we now turn to
effectively one-dimensional harmonic traps.


{\bf A thermal state of a BEC in a 1D harmonic trap.---} In a sufficiently
anisotropic trap $V(x,y,z)=\frac12m\omega^2[x^2+\kappa^2(y^2+z^2)]$ with
$\kappa\gg 1$ the $y-z$ state of all atoms is frozen in the ground state.
The condensate wave function $\phi_0(x)$ solves a 1D Gross-Pitaievskii
equation (\ref{GPE}) with an effective $g_{\rm 1D}=g/\xi^2$, where $\xi$
is the size of the ground state in the $y-z$ plane. We solved the 1D
Bogoliubov-de Gennes equations \cite{BT} to get $u_m$ and $v_m$ for two
sets of parameters relevant to the quasi-1D \cite{Hannover} and strictly
1D \cite{1D} experiments.  Relative depletions $d=dN/N$ corresponding to
the thermal state (\ref{thermal}) are listed in Table~\ref{tab}.
\begin{table}
\caption{\label{tab} Lower bound for the stationary depletion due to atomic 
losses for parameters corresponding to quasi-1D {\protect \cite{Hannover}} 
and 1D {\protect \cite{1D}} experiments. In the calculations the values of 
$g_{1D}N/\hbar\omega$ have been estimated to be 7500 (where $N=1.5\cdot 10^5$) 
and 500 ($N=10^4$), respectively.}
\begin{ruledtabular}
\begin{tabular}{lll}
 & 1-body losses & 3-body losses \\
\hline
quasi-1D   & 6\%   &   10\%   \\
1D         & 2\%   &   4\%   \\
\end{tabular}
\end{ruledtabular}
\end{table}
These values give lower bounds for the stationary depletions in these
experiments. The 10\% depletion in the 3-body losses for the quasi-1D
condensate is actually close to the thermal cloud fraction estimated in
the Hannover experiment \cite{Hannover}. In other words, this experiment
is close to the minimal stationary depletion set by atomic losses.
    

{\bf Relaxation time.---} It takes time to reach the thermal state
(\ref{thermal}). To simplify notation we assume here that losses 
are dominated by only one of the channels $l$. Time evolution of the
average number of quasiparticles 
$n_m(t)={\rm Tr}\hat\rho(t)\hat b_m^{\dagger}\hat b_m$ directly follows 
from the 
master equation (\ref{masterb}) and satisfies a differential equation
\begin{equation}
\frac{d}{dt}n_m(t) = - l\gamma_l\alpha_{lm} N^{l-1}(t) [n_m(t)-n_{lm}]~.        
\label{dn/dt}
\end{equation}
$n_m(t)$ is relaxing toward its equilibrium value $n_{lm}$. Eq.(\ref{dn/dt}) 
has to be compared with the decay law for the total number of atoms 
$N(t)={\rm Tr}\rho(t)\int d^3x~\hat\psi^{\dagger}\hat\psi$
\begin{equation}
\frac{d}{dt}N(t)=-l\gamma_l\alpha_l N^l(t)~.
\label{dN/dt}
\end{equation} 
Here $\alpha_l=\int d^3x~|\phi_0|^{2l}$. This equation is valid for small 
depletion, when almost all atoms occupy $\phi_0(\vec x)$.

As dominant Bogoliubov modes are localized on the condensate, we can
approximate $\alpha_{lm}\approx\alpha_l$. Consequently the relaxation and
decay rates in Eqs.(\ref{dn/dt}) and (\ref{dN/dt}) are comparable,
$dn_m/n_m\approx dN/N$. As the equilibrium value $n_{lm}$ in
Eq.(\ref{dn/dt}) depends on the time-dependent $N(t)$ and the rates are
comparable, $n_m(t)$ will never quite reach the instantaneous equilibrium
value $n_{lm}(t)$. In the following we will consider 1- and 3-body losses
in the two mode (double well) toy model to see that the estimated
equilibrium depletion, we have considered so far, is a {\it lower} bound
for a stationary depletion.


{\bf Double well model.---} The model is described by Hubbard Hamiltonian
\begin{equation}
\hat H_{2}=-\Omega(\hat a_1^{\dagger}\hat a_2+
                   \hat a_2^{\dagger}\hat a_1)+
           \frac12( \hat a_1^{\dagger} \hat a_1^{\dagger} \hat a_1 \hat a_1 +
                    \hat a_2^{\dagger} \hat a_2^{\dagger} \hat a_2 \hat a_2 )~.
\end{equation}
Here we use rescaled dimensionless units such that $\hbar=1$. The master
equation (\ref{master}) becomes
\begin{equation}
\frac{d\hat\rho}{dt}=
\frac{1}{i}[\hat H_2,\hat\rho]+
\gamma_l 
\sum_{j=1,2}
{\cal D}[\hat a_j^l]\rho~.
\label{master2}
\end{equation}
We assume that $l$-body losses dominate.

The condensate wavefunction $\phi_0=(1,1)/\sqrt{2}$ and
$\phi_1=(1,-1)/\sqrt{2}$ span the single particle Hilbert space. There is
only one Bogoliubov mode with $u_1=X\phi_1/\sqrt{X^2-1}$ and
$v_1=-\phi_1/\sqrt{X^2-1}$, where
$X=\left(1+\frac{4\Omega}{N}\right)+
\sqrt{\left(1+\frac{4\Omega}{N}\right)^2-1}$.
As $\phi_0$ is ``uniform", the equilibrium number of quasiparticles is
independent of $l$,
\begin{equation}
n_{l1}=dN_1^{(0)}=
v_1^{\dagger}v_1=
\sqrt{\frac{N}{32\Omega}}-\frac12+{\cal O}(N^{-1/2})~,
\label{nl1}
\end{equation}
Depletion in the equilibrium thermal state (\ref{thermal}) is
\begin{equation}
d=\frac{dN_1^{(0)}+\left(1+2dN_1^{(0)}\right)n_{l1}}{N}=
\frac{1}{16\Omega}+{\cal O}(N^{-1})~.
\label{deq}
\end{equation}


{\bf 1-body losses.---} We begin with $l=1$ and a condensate initially in 
the Bogoliubov vacuum
state, i.e. the number of quasiparticles $n_1(0)=0$. 
A formal solution of Eq.(\ref{dn/dt}), where $\alpha_{l1}=1$, is 
$n_1(t)=\gamma_1\int_0^tdt'~\exp[-\gamma_1(t-t')]n_{11}(t')$.
$N(t)=N_0e^{-\gamma_1t}$, which solves
Eq.(\ref{dN/dt}), and Eqs.(\ref{nl1},\ref{dN}) give 
(time-dependent) depletion
\begin{equation}
d^{l=1}(t)=
\frac{1-\sqrt{f}}{8\Omega}+\frac{f}{\sqrt{32\Omega N}}+{\cal O}(N^{-1})~.
\label{d10}
\end{equation}
Here $f=N(t)/N_0$ is a fraction of atoms remaining in the trap. When $f\to 0$
(the system is forgetting about the initial conditions) the depletion (\ref{d10}) 
becomes roughly twice the equilibrium value (\ref{deq}). 
At small $f$ the system is reaching a stationary state with
twice the equilibrium depletion. 

When we initially prepare the system with
$n_1(0)=2\sqrt{\frac{N_0}{32\Omega}}-\frac12$, instead
of the rather arbitrary $n_1(0)=0$, then the depletion is stationary 
--- it does not depend on $f$
(and through $f$ on the initial $N_0$),
\begin{equation}
d^{l=1}_{\rm stat.}=\frac{1}{8\Omega}+{\cal O}(N^{-1})~.
\label{d1qeq}
\end{equation}
This stationary depletion
(\ref{d1qeq}) is roughly twice the equilibrium value (\ref{deq}). 


{\bf 3-body losses.---} When 3-body losses dominate, the system gets much
closer to the equilibrium than in the case of 1-body losses. For a
condensate initially in the Bogoliubov vacuum state, $n_1(0)=0$, a similar
procedure as for $l=1$ leads to a depletion
\begin{equation}
d^{l=3}(t)=
\frac{3(1-f^{5/2})}{40\Omega}+
\frac{f^3}{\sqrt{32\Omega N}}+{\cal O}(N^{-1})~. 
\label{d30}
\end{equation}
When $f\to 0$ $~d^{l=3}(t)$ is approaching a stationary value 
\begin{equation}
d^{l=3}_{\rm stat.}=\frac{3}{40\Omega}+{\cal O}(N^{-1})~,
\label{d3qeq}
\end{equation}
that is only $1.2$ higher than the equilibrium value (\ref{deq}).  
Starting with the initial
$n_1(0)=\frac65\sqrt{\frac{N_0}{32\Omega}}-\frac12$ results in a depletion
independent of $f$ (and thus also on $N_0$) and equal to the stationary
value (\ref{d3qeq}).

We conclude that our estimates of depletion based on the equilibrium values 
are lower bounds for stationary depletions.

\begin{figure}[htb]
\includegraphics*[width=8.6cm]{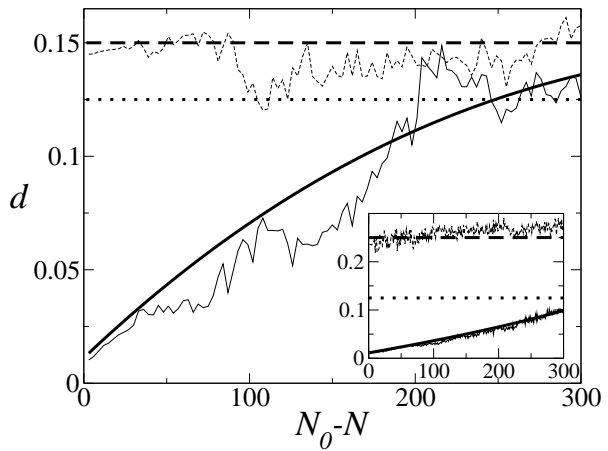}
\caption{ Relative depletion $d=dN/N$ in the 2-mode model, as a function
of the number of atoms that escaped from the trap due to 3-body losses
(for $N_0=500$, $\Omega=0.5$ and $\gamma_3=10^{-6}$). The solid lines
correspond to the condensate initially in the Bogolubov vacuum (thick
line: Eq.(\ref{d30}), thin line: numerical results). Dashed lines show the
predicted stationary value $d^{l=3}_{\rm stat.}$ (thick line:
Eq.(\ref{d3qeq}), thin line: numerical results). The dotted line is the
equilibrium value (\ref{deq}). The inset shows the same as in the main
figure but for 1-body losses with $\gamma_1=10^{-2}$.
}
\label{Fig}        
\end{figure}

{\bf Numerical experiment.---} We verified the predictions 
(\ref{d10},\ref{d1qeq},\ref{d30},\ref{d3qeq}) of the Bogoliubov theory in
numerical simulations. For a large $N_0$ a direct solution of the master 
equation (\ref{master2}) is not the most efficient. 
It is better to replace the deterministic $\rho(t)$ by an ensemble
of stochastic pure states $|\Psi(t)\rangle$, such that
$\rho(t)$ is reproduced as an average over many stochastic realizations,
$\overline{|\Psi(t)\rangle\langle\Psi(t)|}=\rho(t)$. A stochastic
``unraveling'' of the master equation (\ref{master2}) is given by
Ito stochastic nonlinear Schr\"odinger equation \cite{Carmichael}
\begin{eqnarray}
d|\Psi\rangle &=&
-idt \hat H_2|\Psi\rangle-
\frac{dt\gamma_l}{2}
\sum_{j=1,2}
\left[
(\hat a_{j}^{\dagger})^l\hat a_{j}^l-
\left\langle 
(\hat a_{j}^{\dagger})^l\hat a_{j}^l 
\right\rangle
\right]|\Psi\rangle 
\nonumber\\
&&
+\sum_{j=1,2}
d{\cal N}_{j}(t) 
\left[ 
\frac{\hat a_j^l |\Psi\rangle}
{\sqrt{ 
\left\langle
(\hat a_{j}^{\dagger})^l\hat a_{j}^l
\right\rangle 
}}
-|\Psi\rangle \right] ~,
\end{eqnarray}
where $\langle\hat A\rangle\equiv\langle\Psi|\hat A|\Psi\rangle$.  
$d{\cal N}_{j}(t)\in\{0,1\}$ is a stochastic process [$d{\cal N}_{j}(t)=1$
when $l$ atoms escape from a well $j$ between $t$ and $t+dt$, and $0$
otherwise]. A probability that $l$ atoms will escape between $t$ and
$t+dt$ is $dt\langle(\hat a_{j}^{\dagger})^l\hat a_{j}^l\rangle$.
In Fig.\ref{Fig} we compare the predictions
(\ref{d10},\ref{d1qeq},\ref{d30},\ref{d3qeq}) with corresponding averages
over many stochastic realizations. As they compare quite well, the
Bogoliubov theory (\ref{masterb}) passes the test on the double-well
model.


{\bf Conclusion.---} A condensate is heated by atomic losses. The
depletion of the system approaches a stationary value that ranges from
around $1\%$ for a uniform 3D condensate to around $10\% $ for a quasi-1D
harmonic trap. As atomic losses cannot be easily eliminated, this
depletion is a serious limitation on quantum coherence of atomic BEC. We
only note here that outcoupling in the atom laser is a non-markovian
process of atomic losses. Its influence on laser coherence will be
addressed elsewhere.


{\bf Acknowledgements.---} We are grateful to Zbyszek Karkuszewski for
discussions. This research was supported in part by KBN grants 2 P03B 092
23 (JD) and 5 P03B 088 21 (KS).


\end{document}